\documentclass[oldversion]{aa}
\usepackage{txfonts}
\usepackage{graphicx}
\usepackage{natbib}
\usepackage{amssymb}
\bibpunct{(}{)}{;}{a}{}{,}
\begin{document}
   \title{Low-mass star formation in R Coronae Australis: Observations of organic molecules with the APEX telescope}
 
   \titlerunning{Low-mass starformation in Coronae Australis}

   \author{F. L. Sch\"oier\inst{1} \and
             J.~K. J{\o}rgensen\inst{2} \and
	   K.~M. Pontoppidan\inst{3} \and
	   A.~A. Lundgren\inst{4}}

   \offprints{F. L. Sch\"oier \\ \email{fredrik@astro.su.se}}

   \institute{Stockholm Observatory, AlbaNova University Center, SE-106 91 Stockholm, Sweden
     \and
	     Harvard-Smithsonian Center for Astrophysics, 60 Garden Street MS42, Cambridge, MA 02138, USA
             \and
	     Division of Geological and Planetary Sciences, California Institute of Technology, MS
                     150-21, Pasadena, CA 91125, USA
                     \and
                      European Southern Observatory, Casilla 19001, Santiago 19, Chile}    
             
   \date{Received; accepted}

   \abstract{This paper presents new APEX submillimetre molecular line observations of three low-mass protostars, IRS7A, IRS7B, and IRAS32, in the R~Coronae~Australis molecular cloud complex. The molecular excitation analysis is performed using a statistical equilibrium radiative transfer code. The derived beam averaged fractional abundances vary by less than a factor of two among the three sources, except those of H$_2$CO and CH$_3$OH, which show differences of about an order of magnitude.  The molecular abundances  are similar to those typically found in other star-forming regions in the Galaxy, such as the $\rho$~Oph and Perseus molecular clouds. There is a marked difference in the kinetic temperatures derived for the protobinary source IRS7 from H$_2$CO ($\approx$\,40--60\,K) and CH$_3$OH ($\approx$\,20\,K), possibly indicating a difference in origin of the emission from these two molecules.
   
   }
   \maketitle
%

%
   \begin{figure*}
   \centering{   
   \includegraphics[width=11.7cm]{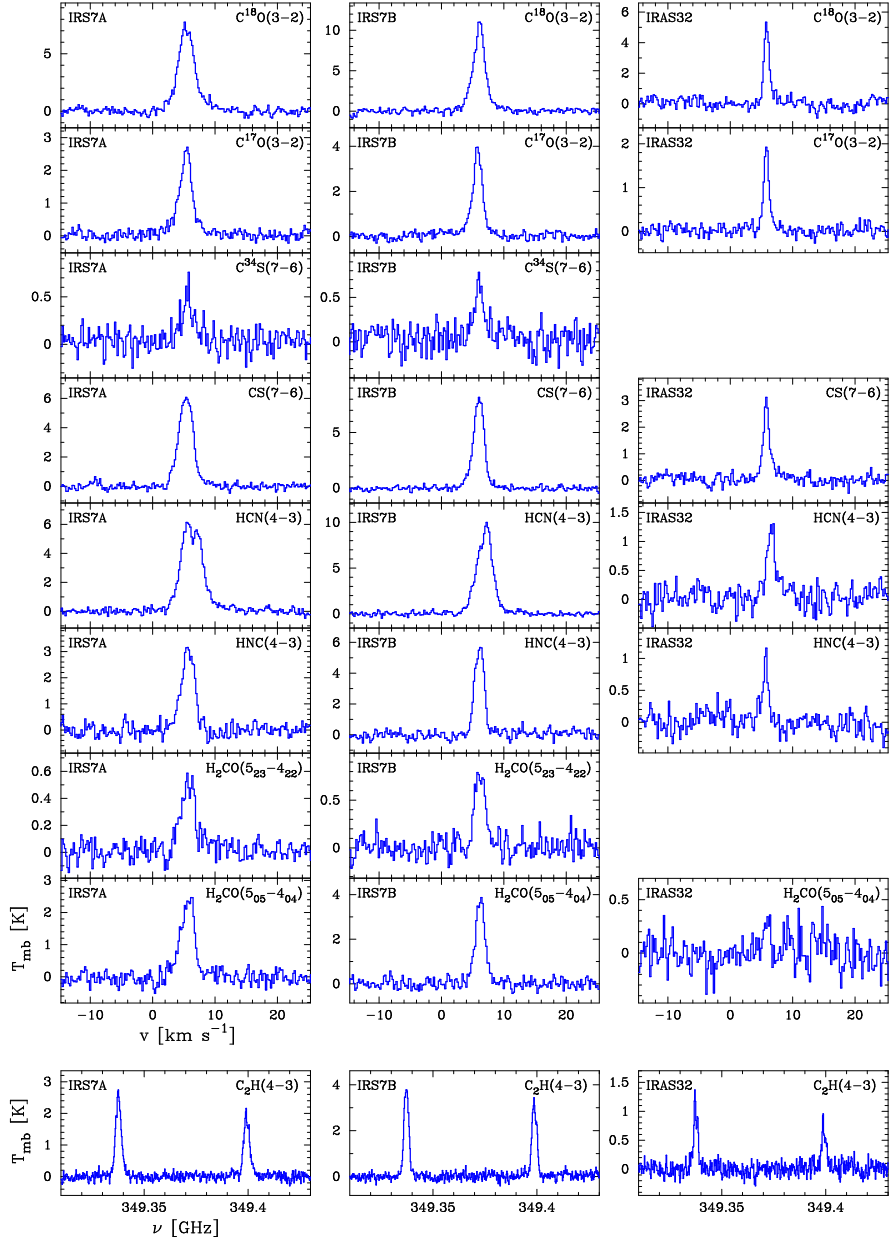}
   \caption{New observations of molecular line emission  using the APEX telescope. The velocity resolution is 0.2\,km\,s$^{-1}$ in all the spectra.}
   \label{obs_apex}}
   \end{figure*}

\section{Introduction}
In the earliest, deeply-embedded stage, low-mass protostars are
 surrounded by a collapsing envelope and a circumstellar disk.
In this phase, 
the initial chemical composition of the disk is determined.  The chemistry of molecules in the circumstellar envelope 
is likely to be directly reflected in the molecular composition in the disk, which forms the building blocks from which planets may form.
An interesting aspect is the search for  low-mass
'hot cores`, i.e., regions where grain ice mantles are liberated in
the envelope close to the central protostar because of its heating
\citep[e.g.,][]{Ceccarelli00b, Schoeier02a, Cazaux03}. The search for hot cores around low-mass stars is not yet
conclusive, mainly due to limitations in the available observational
data. Also, outflows are likely to play an important role by affecting many of the same species thought to be sign posts of hot cores
\citep[e.g.,][]{Joergensen05b}. Moreover, the chemistry of traditional hot-core tracers such as
H$_2$CO may be affected by the freeze-out of CO \citep[e.g.,][]{Schoeier04a}, which is why combined studies of chemically related molecular species are important.

The R Coronae Australis dark cloud is a nearby \citep[170\,pc,][]{Knude98} region of active star formation \citep[e.g.,][]{Brown87, Chini03, Nutter05}.
Molecular line observations of protostellar objects in R Coronae Australis  have attained little attention, mainly due to its location on the southern hemisphere. 
Previous studies using single-dish telescopes with limited spatial resolution, which are mainly sensitive
to the chemistry of the outer cold regions of the envelopes, have been biased towards lower excitation lines. A wealth of molecular rotational transitions lying in the 279\,$-$\,381\,GHz window are accessible with APEX\footnote{This publication is based on data acquired with the Atacama Pathfinder Experiment (APEX). APEX is a collaboration between the Max-Planck-Institut f\"ur Radioastronomie, the European Southern Observatory, and the Onsala Space Observatory}, and can be used to constrain the 
physics, and in particular, the chemistry in the dense
($>$\,10$^{6}$\,cm$^3$) and warm ($>$\,30\,K)
material in the envelopes.
Observations of protostellar objects in Corona Australis are also interesting for testing the importance of the
surroundings on the formation of protostars, and will complement
 previous extensive studies in Perseus and $\rho$ Oph \citep{Joergensen02, Joergensen04c, Joergensen05b}. It is not clear how the
properties of the parental cloud affect 
the properties of
the emerging protostar \citep[see][]{Jayawardhana01}.
Likewise, the properties of the parental cloud may
reflect the chemical evolution of the protostar, for example, in the
degree of freeze-out and the 'initial abundances`.

The first step in a project aimed at obtaining an extensive data base of molecular line emission towards a large sample of protostars in the southern hemisphere, with a focus on the Corona Australis and Chameleon molecular clouds, is presented in this {\em Letter}. 
It is crucial to utilize APEX to build comparable samples of well-studied low-mass protostars for future ALMA observations

\section{Observations}
\label{sect_obs}
Observations of sub-millimetre molecular line emission 
towards the R Coronae Australis molecular cloud complex were performed September 4--6, 2005, using the 12\,m APEX telescope equipped with an SIS receiver (APEX--2a). Typical system temperatures were in the range 140\,$-$\,380\,K (DSB). The 1\,GHz FFTS with 8192 spectral channels was used as a backend, providing a spectral resolution of 122\,kHz (0.1 km\,s$^{-1}$). The observations were carried out using a position-switching mode, with the reference position located at +2$\arcmin$ in azimuth. 
The data were reduced in a standard way by removing, in most cases, a low order ($\leq$\,3) polynomial baseline and the raw spectra, stored in the $T_{\mathrm A}^{\ast}$  scale, converted to main-beam brightness temperature using $T_{\mathrm{mb}}$\,=\,$T_{\mathrm
A}^{\ast}/\eta_{\mathrm{mb}}$. $T_{\mathrm A}^{\ast}$ is the
antenna temperature corrected for atmospheric attenuation using the
chopper-wheel method, and $\eta_{\mathrm{mb}}$ is the main-beam
efficiency. A value of $\eta_{\mathrm{mb}}$\,$=$\,0.7 was assumed for all spectral settings.
Regular pointing checks were made on strong CO sources and typically found to be consistent within $\approx$\,3$\arcsec$. The FWHM of the main beam is 18\arcsec at 345\,GHz.
The uncertainty (including errors in pointing, calibration, and $\eta_{\mathrm{mb}}$) in the absolute intensity scale (main-beam brightness) is estimated to be about $\pm 15$\%.

\begin{figure}
   \centering{   
   \includegraphics[width=6.3cm]{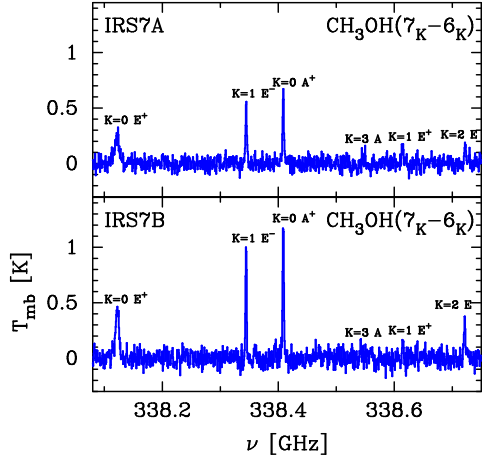}
   \caption{New observations of CH$_3$OH line emission using the APEX telescope. The velocity resolution is 0.5\,km\,s$^{-1}$ in both spectra. Note that the $7_0$\,$\rightarrow$\,$6_{0}$ E$^{+}$ line is blended with the CH$_3$OH $4_0$\,$\rightarrow$\,$3_{1}$ E$^{-}$ line from the upper sideband.}
   \label{obs_ch3oh}}
   \end{figure}

The observed spectra for the protostellar objects IRS7A ($\alpha_{2000}$\,=\,19:01:55.3, $\delta_{2000}$\,=\,--36:57:21), IRS7B ($\alpha_{2000}$\,=\,19:01:56.4, $\delta_{2000}$\,=\,--36:57:27), and IRAS32 ($\alpha_{2000}$\,=\,19:02:58.7, $\delta_{2000}$\,=\,--37:07:34) are presented in Figs.~\ref{obs_apex} and \ref{obs_ch3oh} (the positions are obtained from IRAC4 8\,$\mu$m pipeline data in the Spitzer Space Telescope archive: AOR ID 3650816). IRS7B is identified as a class 0 protostar, whereas  IRS7A and IRAS32 are of class I \citep[see][ and references therein]{Nutter05}.

Results of Gaussian fits to the spectra are reported in Table~\ref{intensities}. The systemic (LSR) velocities obtained from the Gaussian fits are 5.3, 5.6, and 5.9\,km\,s$^{-1}$ (LSR) for IRS7A, IRS7B, and IRAS32, respectively. The spectra are generally well represented by a single Gaussian, with the major exception being the observed HCN $J$\,=\,4\,$\rightarrow$\,3 line emission towards IRS7A and IRS7B. In these two cases, not only is the HCN line significantly broader, compared to other molecular lines, but also a central absorption dip is present, at least in the IRS7A spectrum. The HCN spectra presented here 
are very similar to those of HCO$^{+}$ $J$\,=\,4\,$\rightarrow$\,3 obtained by \citet{Anderson97b}. They interpret the emission as arising in a large rotating (circumbinary) disk, and the self-absorption feature is caused by absorption in a dense and quiescent foreground of the cloud. In addition, for the two IRS7 sources,  our observed C$^{17}$O and, in particular, C$^{18}$O,  $J$\,=\,3\,$\rightarrow$\,2 spectra have weak wings, possibly related to the known molecular outflow in this region, and are somewhat better reproduced by a sum of two Gaussians.

\begin{table}
\caption{Observational results$^{\mathrm{a}}$.}
\label{intensities}
$
\begin{array}{p{0.2\linewidth}ccccccccccc}
\hline
\noalign{\smallskip}
&
\multicolumn{3}{c}{\mathrm{IRS7A}} &&
\multicolumn{3}{c}{\mathrm{IRS7B}} &&
\multicolumn{3}{c}{\mathrm{IRAS\ 32}} 
\\

\cline{2-4}
\cline{6-8}
\cline{10-12}
\multicolumn{1}{c}{\mathrm{Transition}} &
\multicolumn{1}{c}{I} &
\multicolumn{1}{c}{T_{\mathrm{mb}}} &
\multicolumn{1}{c}{\Delta v} &&
\multicolumn{1}{c}{I} &
\multicolumn{1}{c}{T_{\mathrm{mb}}} &
\multicolumn{1}{c}{\Delta v} &&
\multicolumn{1}{c}{I} &
\multicolumn{1}{c}{T_{\mathrm{mb}}} &
\multicolumn{1}{c}{\Delta v}
\\
\noalign{\smallskip}
\hline
\noalign{\smallskip}
C$^{18}$O 3--2                            & 24.9 & 7.2$:$   & 3.1$:$  && 26.9 & 10.2$:$ & 2.3$:$ && 6.5 & 5.2 & 1.0 \\
C$^{17}$O 3--2                            &   7.5 &  2.5$:$   & 2.5$:$  &&   8.8 &    3.7$:$ & 1.9$:$ && 2.6 & 1.9 & 1.0 \\
C$^{34}$S 7--6                            &   1.4 &  0.47 & 2.5 &&   2.0 &    0.61 & 2.1 && {<}0.2 & \cdots & \cdots \\
CS 7--6                                          &  17.6 &  6.2 & 2.7 &&  17.6 &  8.1 & 2.0 && 4.6 & 2.8$:$ & 1.3$:$ \\
HCN 4--3                                       &  26.0 &  6.1$:$ & 3.9$:$ &&  30.4 &  9.2$:$ & 3.0$:$ && 2.7 & 1.2$:$ & 1.8$:$ \\
HNC 4--3                                       &  8.1 &  3.1 & 2.4 &&  11.3 &  5.9 & 1.8 && 1.8 & 1.1$:$ & 1.0$:$ \\
C$_2$H:                                        &\\
  \ 4$_5$--3$_4$                            &   7.7 &  2.6 & 2.8 &&   10.0 &  4.0 & 2.3 && 2.4 & 1.1$:$ & 2.1$:$ \\
  \ 4$_4$--3$_3$                           &   6.4 &  1.9 & 2.9 &&   8.4 &    2.2 & 2.4 && 2.1 & 0.7$:$ & 2.4$:$ \\
H$_2$CO:                                     &\\
  \ 5$_{23}$--4$_{22}$                  &   1.8 &  0.51 & 2.8 &&   1.7 &    0.81 & 2.0 && {<}0.2 & \cdots & \cdots \\
  \ 5$_{05}$--4$_{04}$                  &   6.2 &  2.4 & 2.5 &&   7.7 &    3.8 & 1.8 && 0.3$:$ & \cdots & \cdots \\
CH$_3$OH:                                 & \\
  \ 7$_0$--6$_0$ E$^+$$^{\mathrm b}$        &  2.3 & \cdots & \cdots &&  3.0 & \cdots & \cdots && {<}0.1 & \cdots & \cdots \\
  \ 7$_1$--6$_1$ E$^-$         & 1.5 & 0.50 & 2.6 &&  2.2 &  1.0  & 2.1 && {<}0.1 & \cdots & \cdots  \\
  \ 7$_1$--6$_1$ E$^+$        & 0.49 & \cdots & \cdots &&  0.37 & \cdots & \cdots &&  {<}0.1 & \cdots & \cdots \\
  \ 7$_2$--6$_2$ E$^{\pm}$ &  0.51 & 0.18$:$  & 3.0$:$ && 0.90 & 0.35$:$ & 2.6$:$  &&  {<}0.1 & \cdots & \cdots \\
 \ 7$_0$--6$_0$ A$^+$          & 1.8 & 0.64 & 2.4 && 3.3 & 1.2 & 2.3 &&  {<}0.1 & \cdots & \cdots \\
  \ 7$_3$--6$_3$ A$^{\pm}$  & 0.2$:$  & \cdots & \cdots &&  0.2$:$ & \cdots & \cdots && {<}0.1 & \cdots & \cdots  \\
\noalign{\smallskip}
\hline
\end{array}
$

$^{\mathrm a}$ $I$\,$=$\,$\int T_{\mathrm{mb}}\,dv$ is the integrated intensity in K\,km\,s$^{-1}$. $T_{\mathrm{mb}}$ is the peak intensity (in K) and $\Delta v$ the FWHM (in km\,s$^{-1}$) from Gaussian fits. A colon (:) marks an uncertain value due to low a S/N-ratio or a significant departure from a Gaussian line profile.

$^{\mathrm b}$ Blended with the CH$_3$OH 4$_0$\,$\rightarrow$\,3$_1$ E$^-$ line from the upper sideband. The integrated intensity refers to the sum of both lines. 
\end{table}

\section{Excitation analysis}
\label{sect_model}

Presently, there is a lack of observational data required to carry out a detailed modelling of the dust emission,  mainly due to limited spatial resolution \citep[see][]{Nutter05}. This is usually the preferred method to determine the physical structure, such as density and temperature, of material present around protostars that is required for a detailed analysis of its molecular line emission  \citep[e.g.,][]{Joergensen02}.

A first estimate of the excitation conditions can be obtained assuming the lines to be emanating from the same homogeneous region. We have adopted RADEX, a statistical equilibrium radiative transfer code that uses an escape probability formalism to calculate the excitation and emission from the protostellar cores in our sample. RADEX is
comparable to the LVG method and provides a useful tool for rapidly
analysing a large set of observational data providing constraints on
physical conditions, such as density and kinetic temperature (Jansen et al.\ 1994\nocite{Jansen94}; van der Tak et al., in prep.).
An online version of RADEX, together with relevant molecular data used in the excitation analysis and summarized in \citet{Schoeier05a},  is made publicly available through the {\em Leiden Atomic and Molecular Database} (LAMDA){\footnote{\tt http://www.strw.leidenuniv.nl/$\sim$moldata}}.




In the excitation analysis, the number density of H$_2$ molecules was fixed to 3\,$\times$\,10$^6$\,cm$^{-3}$, typical of the warmer, more dense regions of the condensed cores. From the two lines of H$_2$CO, the kinetic temperatures were estimated to be $\approx$\,60\,K and $\approx$\,40\,K for IRS7A and IRS7B, respectively. The corresponding column densities derived were 2\,$\times$\,10$^{13}$\,cm$^{-2}$ (IRS7A) and  3.5\,$\times$\,10$^{13}$\,cm$^{-2}$ (IRS7B). In the case of CH$_3$OH, a full analysis using a $\chi^2$-statistic was used, and the kinetic temperatures derived were 23$\pm$3\,K and  22$\pm$3\,K for IRS7A and IRS7B, respectively. The column densities of CH$_3$OH derived were 1.4$\pm$0.4\,$\times$\,10$^{14}$\,cm$^{-2}$ (IRS7A) and 1.8$\pm$0.4\,$\times$\,10$^{14}$\,cm$^{-2}$ (IRS7B). The substantial difference in the derived kinetic temperatures from H$_2$CO and CH$_3$OH could possibly indicate different origins for these traditional hot-core tracers.

To determine the column densities for the remaining molecular species, we adopted a kinetic temperature of 40\,K. The column densities obtained for C$^{34}$S in the case of IRS7A (1\,$\times$\,10$^{13}$\,cm$^{-2}$) and IRS7B (1.5\,$\times$\,10$^{13}$\,cm$^{-2}$) were used together with a 'standard` fractional abundance (5\,$\times$\,10$^{-11}$) of C$^{34}$S \citep[][ see Table~\ref{abundances}]{Joergensen04c} to estimate the total H$_2$ column densities. In the case of IRS7A, we obtain a H$_2$ column density of 2\,$\times$\,10$^{23}$\,cm$^{-2}$, and for IRS7B, a slightly higher value of 3\,$\times$\,10$^{23}$\,cm$^{-2}$.
 For IRAS32, no C$^{34}$S $J$\,=\,7\,$\rightarrow$\,6 line emission was detected, and instead the main isotope (with an adopted fractional abundance of 1\,$\times$\,10$^{-9}$)  was used to estimate the H$_2$ column density to 3\,$\times$\,10$^{22}$\,cm$^{-2}$. The derived fractional abundances for all molecular species observed are reported in Table~\ref{abundances}.

\begin{table}
\caption{Model results.}
\label{abundances}
\resizebox{\hsize}{!}{
$ 
\begin{array}{p{0.16\linewidth}ccccccc}
\hline
\noalign{\smallskip}
&
\multicolumn{1}{c}{\mathrm{IRS7A}} &&
\multicolumn{1}{c}{\mathrm{IRS7B}} &&
\multicolumn{1}{c}{\mathrm{IRAS32}} &&
\multicolumn{1}{c}{\mathrm{standard}^{a}} 
\\
\cline{2-2}
\cline{4-4}
\cline{6-6}
\cline{8-8}
\multicolumn{1}{c}{\mathrm{Molecule}} &
\multicolumn{1}{c}{f(\mathrm{X})/f(\mathrm{H}_2)} &&
\multicolumn{1}{c}{f(\mathrm{X})/f(\mathrm{H}_2)} &&
\multicolumn{1}{c}{f(\mathrm{X})/f(\mathrm{H}_2)} &&
\multicolumn{1}{c}{f(\mathrm{X})/f(\mathrm{H}_2)}
\\
\noalign{\smallskip}
\hline
\noalign{\smallskip}
C$^{18}$O                             & 6.0\times10^{-8}\phantom{:^{1}}          && 5.0\times10^{-8}\phantom{:^{1}}           && \phantom{<}1.0\times10^{-7}\phantom{:^{1}}          && 4.1\times10^{-8}\phantom{^{1}}  \\
C$^{17}$O                             & 2.0\times10^{-8} \phantom{:^{1}}         &&  1.5\times10^{-8}\phantom{:^{1}}          && \phantom{<}2.5\times10^{-8}\phantom{:^{1}}          && 1.4\times10^{-8}\phantom{^{1}}  \\
C$^{34}$S                             &  5.0\times10^{-11}$:$                             && 5.0\times10^{-11}$:$                              && \phantom{<}\cdots\phantom{:^{1}}                            &&  5.1\times10^{-11} \\
CS                                           &  8.0\times10^{-10}\phantom{:}               && 6.0\times10^{-10}\phantom{:}           && \phantom{<}1.0\times10^{-9}$:$\phantom{^{1}}    && 2.5\times10^{-9} \phantom{^{1}} \\
C$_2$H                                  & 5.0\times10^{-10}\phantom{:}           &&  4.0\times10^{-10}\phantom{:}           &&  \phantom{<}9.0\times10^{-10}\phantom{:}        && \cdots \phantom{:^{1}} \\
HCN                                        &  \cdots\phantom{:^{1}}                             && \cdots\phantom{:^{1}}                               && \phantom{<}\cdots \phantom{:^{1}}                          && 2.2\times10^{-9} \phantom{^{1}}\\
HNC                                        &  1.0\times10^{-10}\phantom{:}                && 1.0\times10^{-10}\phantom{:}          && \phantom{<}1.2\times10^{-10}\phantom{:}       && 2.5\times10^{-10}\\
p-H$_2$CO                           &  2.0\times10^{-10}\phantom{:}                && 1.0\times10^{-10}\phantom{:}          && \phantom{<}4.0\times10^{-11}\phantom{:}       && 6.9\times10^{-10}\\
A-CH$_3$OH                        & 6.0\times10^{-10}\phantom{:}           && 7.0\times10^{-10}\phantom{:}            && {<}7.0\times10^{-11}\phantom{:}                           && 1.8\times10^{-9}\phantom{^{1}} \\
E-CH$_3$OH                        & 7.0\times10^{-10}\phantom{:}           && 6.0\times10^{-10}\phantom{:}           && {<}1.3\times10^{-10}\phantom{:}                            && 1.3\times10^{-9}\phantom{^{1}} \\
\noalign{\smallskip}
\hline
\end{array}
$}
$^a$ Median values from  \citet{Joergensen02}, \citet{Joergensen04c}, and \citet{Joergensen05b}. Note that CH$_3$OH was only detected in known outflow sources.\\
A colon (:) marks an adopted value used to set the density scale.
\end{table}

\section{Results and discussion}
The excitation analysis gives results (see Table~\ref{abundances}) for the abundances of common molecules such as CO, CS, and HNC, and for more complex species, such as H$_2$CO and CH$_3$OH, that are consistent with results from other star-forming regions, such as $\rho$~Oph and Perseus \citep{Joergensen02, Joergensen04c, Joergensen05b}. The derived fractional abundances vary by less than a factor of two among the three sources for the majority of molecular species. In the present analysis, it is not possible to correlate the abundances with the identified protostellar class.
It is interesting to note that the abundances of both H$_2$CO and CH$_3$OH vary by about an order of magnitude among the three sample sources, possibly indicating a difference in central heating source or impact of outflows. We find the abundances of these molecules to be particularly low in IRAS32.

IRS7A and IRS7B form a protobinary system with a separation of about 18$\arcsec$ (3000\,AU). As such,  RCrA--IRS7 is also a cousin of the well-studied IRAS 16293--2422 and NGC--IRAS4 binaries in terms of the total luminosity. The fact that \citet{Joergensen05b} only detected CH$_3$OH emission from known outflow sources in their large survey of protostellar cores, together with the low kinetic temperature ($\approx$\,20\,K) we derive from our observed CH$_3$OH emission in contrast to what we get for H$_2$CO ($\approx$\,40--60\,K), suggests the impact of outflows in regulating the chemistry around IRS7. These temperatures are similar to those obtained for NGC--IRAS4 \citep{Maret04,Maret05}, but lower than those for IRAS 16293--2422 \citep{Dishoeck95}. However, in contrast to the NGC--IRAS4A binary in particular, IRS7 does not show CH$_3$OH lines that are broader than other observed molecular lines. Nonetheless, a large scale outflow thought to be associated with IRS7 has been observed in several other molecular species \citep{Levreault88, Anderson97a}.





To make further progress, in particular to solidify the claims regarding the chemistry in the direct neighbourhood of the observed protostars, more detailed knowledge of the underlying physical structure is required. Such work is in progress and involves mid-IR observations from large ground-based telescopes and the Spitzer space telescope, and interferometric mm/sub-mm observations from ATCA and the SMA. 


\begin{acknowledgements}
We would like to thank the anonymous referee who helped us improve the presentation of this work.
The authors are grateful to the staff at the APEX telescope. 
FLS acknowledges financial support from the Swedish Research Council.
The research of JKJ was supported by the NASA Origins Grant NAG5--13050. 
\end{acknowledgements}

\bibliographystyle{aa}


\begin{thebibliography}{20}
\expandafter\ifx\csname natexlab\endcsname\relax\def\natexlab#1{#1}\fi

\bibitem[{{Anderson} {et~al.}(1997{\natexlab{a}}){Anderson}, {Harju}, \&
  {Haikala}}]{Anderson97b}
{Anderson}, I.~M., {Harju}, J., \& {Haikala}, L.~K. 1997{\natexlab{a}}, \aap,
  326, 366

\bibitem[{{Anderson} {et~al.}(1997{\natexlab{b}}){Anderson}, {Harju}, {Knee},
  \& {Haikala}}]{Anderson97a}
{Anderson}, I.~M., {Harju}, J., {Knee}, L.~B.~G., \& {Haikala}, L.~K.
  1997{\natexlab{b}}, \aap, 321, 575

\bibitem[{{Brown}(1987)}]{Brown87}
{Brown}, A. 1987, \apjl, 322, L31

\bibitem[{{Cazaux} {et~al.}(2003){Cazaux}, {Tielens}, {Ceccarelli}, {Castets},
  {Wakelam}, {Caux}, {Parise}, \& {Teyssier}}]{Cazaux03}
{Cazaux}, S., {Tielens}, A.~G.~G.~M., {Ceccarelli}, C., {et~al.} 2003, \apjl,
  593, L51

\bibitem[{{Ceccarelli} {et~al.}(2000){Ceccarelli}, {Loinard}, {Castets},
  {Tielens}, \& {Caux}}]{Ceccarelli00b}
{Ceccarelli}, C., {Loinard}, L., {Castets}, A., {Tielens}, A.~G.~G.~M., \&
  {Caux}, E. 2000, A\&A, 357, L9

\bibitem[{{Chini} {et~al.}(2003){Chini}, {K{\"a}mpgen}, {Reipurth}, {Albrecht},
  {Kreysa}, {Lemke}, {Nielbock}, {Reichertz}, {Sievers}, \& {Zylka}}]{Chini03}
{Chini}, R., {K{\"a}mpgen}, K., {Reipurth}, B., {et~al.} 2003, \aap, 409, 235

\bibitem[{{Jansen} {et~al.}(1994){Jansen}, {van Dishoeck}, \&
  {Black}}]{Jansen94}
{Jansen}, D.~J., {van Dishoeck}, E.~F., \& {Black}, J.~H. 1994, \aap, 282, 605

\bibitem[{{Jayawardhana} {et~al.}(2001){Jayawardhana}, {Hartmann}, \&
  {Calvet}}]{Jayawardhana01}
{Jayawardhana}, R., {Hartmann}, L., \& {Calvet}, N. 2001, \apj, 548, 310

\bibitem[{{J{\o}rgensen} {et~al.}(2002){J{\o}rgensen}, {Sch{\" o}ier}, \& {van
  Dishoeck}}]{Joergensen02}
{J{\o}rgensen}, J.~K., {Sch{\" o}ier}, F.~L., \& {van Dishoeck}, E.~F. 2002,
  \aap, 389, 908

\bibitem[{{J{\o}rgensen} {et~al.}(2004){J{\o}rgensen}, {Sch{\" o}ier}, \& {van
  Dishoeck}}]{Joergensen04c}
{J{\o}rgensen}, J.~K., {Sch{\" o}ier}, F.~L., \& {van Dishoeck}, E.~F. 2004,
  \aap, 416, 603

\bibitem[{{J{\o}rgensen} {et~al.}(2005){J{\o}rgensen}, {Sch{\"o}ier}, \& {van
  Dishoeck}}]{Joergensen05b}
{J{\o}rgensen}, J.~K., {Sch{\"o}ier}, F.~L., \& {van Dishoeck}, E.~F. 2005,
  \aap, 437, 501

\bibitem[{{Knude} \& {H{\o}g}(1998)}]{Knude98}
{Knude}, J. \& {H{\o}g}, E. 1998, \aap, 338, 897

\bibitem[{{Levreault}(1988)}]{Levreault88}
{Levreault}, R.~M. 1988, \apjs, 67, 283

\bibitem[{{Maret} {et~al.}(2004){Maret}, {Ceccarelli}, {Caux}, {Tielens}, {J{\"
  o}rgensen}, {van Dishoeck}, {Bacmann}, {Castets}, {Lefloch}, {Loinard},
  {Parise}, \& {Sch{\" o}ier}}]{Maret04}
{Maret}, S., {Ceccarelli}, C., {Caux}, E., {et~al.} 2004, \aap, 416, 577

\bibitem[{{Maret} {et~al.}(2005){Maret}, {Ceccarelli}, {Tielens}, {Caux},
  {Lefloch}, {Faure}, {Castets}, \& {Flower}}]{Maret05}
{Maret}, S., {Ceccarelli}, C., {Tielens}, A.~G.~G.~M., {et~al.} 2005, \aap,
  442, 527

\bibitem[{{Nutter} {et~al.}(2005){Nutter}, {Ward-Thompson}, \&
  {Andr{\'e}}}]{Nutter05}
{Nutter}, D.~J., {Ward-Thompson}, D., \& {Andr{\'e}}, P. 2005, \mnras, 357, 975

\bibitem[{{Sch{\" o}ier} {et~al.}(2002){Sch{\" o}ier}, {J{\o}rgensen}, {van
  Dishoeck}, \& {Blake}}]{Schoeier02a}
{Sch{\" o}ier}, F.~L., {J{\o}rgensen}, J.~K., {van Dishoeck}, E.~F., \&
  {Blake}, G.~A. 2002, \aap, 390, 1001

\bibitem[{{Sch{\" o}ier} {et~al.}(2004){Sch{\" o}ier}, {J{\o}rgensen}, {van
  Dishoeck}, \& {Blake}}]{Schoeier04a}
{Sch{\" o}ier}, F.~L., {J{\o}rgensen}, J.~K., {van Dishoeck}, E.~F., \&
  {Blake}, G.~A. 2004, \aap, 418, 185

\bibitem[{{Sch{\" o}ier} {et~al.}(2005){Sch{\" o}ier}, {van der Tak}, {van
  Dishoeck}, \& {Black}}]{Schoeier05a}
{Sch{\" o}ier}, F.~L., {van der Tak}, F.~F.~S., {van Dishoeck}, E.~F., \&
  {Black}, J.~H. 2005, \aap, 432, 369

\bibitem[{{van Dishoeck} {et~al.}(1995){van Dishoeck}, {Blake}, {Jansen}, \&
  {Groesbeck}}]{Dishoeck95}
{van Dishoeck}, E.~F., {Blake}, G.~A., {Jansen}, D.~J., \& {Groesbeck}, T.~D.
  1995, ApJ, 447, 760

\end{thebibliography}

\end{document}